\documentclass[11pt,authoryear,nofootinbib,preprint]{article}
\usepackage{graphicx}
\usepackage{epstopdf}
\usepackage{bm}
\usepackage{latexsym}
\usepackage[square]{natbib}

\usepackage{rotating}

\newcommand{\md}{\mathrm{d}}
\newcommand{\e}{\mathrm{e}}
\newcommand{\w}{\omega}

\begin{document}
\title{What drives mutual fund asset concentration? }
\author{Yonathan Schwarzkopf\footnote{%
California Institute of Technology, Pasadena, CA 91125. Santa Fe Institute, Santa Fe, NM 87501. yoni@caltech.edu.}  %
  \,\,and J. Doyne Farmer\footnote{%
Santa Fe Institute, Santa Fe, NM 87501.Luiss Guido Carli, ROMA Italy. jdf@santafe.edu.}%
}
\date{\today}

\maketitle

\begin{abstract}
Is the large influence that mutual funds assert on the U.S. financial system spread across many funds, or is it is concentrated in only a few?  We argue that the dominant economic factor that determines this is market  efficiency, which dictates that fund performance is size independent and fund growth is essentially random.  The random process is characterized by entry, exit and growth.  We present a new time-dependent solution for the standard equations used in the industrial organization literature and show that relaxation to the steady-state solution is extremely slow.  Thus, even if these processes were stationary (which they are not), the steady-state solution, which is a very heavy-tailed power law, is not relevant.  The distribution is instead well-approximated by a less heavy-tailed log-normal.  
We perform an empirical analysis of the growth of mutual funds, propose a new, more accurate size-dependent model, and show that it makes a good prediction of the empirically observed size distribution.  While mutual funds are in many respects like other firms, market efficiency introduces effects that make their growth process distinctly different.  Our work shows that a simple model based on market efficiency provides a good explanation of the concentration of assets, suggesting that other effects, such as transaction costs or the behavioral aspects of investor choice, play a smaller role.
\end{abstract}

\newpage
\section {Introduction}\label{section_introduction}

In the past decade the mutual fund industry has grown rapidly, moving from $3\%$ of taxable household financial assets in 1980, to $8\%$ in 1990, to $23\%$ in 2007\footnote{
Data is taken from the Investment Company Institute's 2007 fact book available at www.ici.org. }. 
In absolute terms, in 2007 this corresponded to 4.4 trillion USD and 24\% of U.S. corporate equity holdings.  Mutual funds account for a significant fraction of trading volume in financial markets and have a substantial influence on prices.   This raises the question of who has this influence:  Are mutual fund investments concentrated in a few dominant large funds, or spread across many funds of similar size? Do we need to worry that a few funds might become so large that they are ``too big to fail"?  What are the economic mechanisms that determine the concentration of investment capital in mutual funds?

Large institutional investors are known to play an important role in the market \citep{corsetti-2001}. Gabaix et al. recently hypothesized that the fund size distribution plays a central role in explaining the heavy tails in the distribution of both trading volume and price returns\footnote{
The equity fund size distribution was argued to be responsible for the observed distribution of trading volume \citep{levy-1996,solomon-2001}, and Gabaix et al. have argued that it is important for explaining the distribution of price returns \citep{gabaix-2003-nature,gabaix-2006}.}.
If their theory is true this would imply that the heavy tails in the distribution of mutual fund size play an important role in determining market risk.

While it is standard in economics to describe distributional inequalities in terms of statistics such as the Gini or Herfindahl indices, as we show in Appendix A, this approach is inadequate to describe the concentration in the tail.  Instead, the best way to describe the concentration of assets is in terms of the functional form of the tail.  As is well-known in extreme value theory \citep{Embrechts97}, the key distinction is whether all the moments of the distribution are finite.  If the tail is truly concentrated, the tail is a power law, and all the moments above a given threshold, called the tail exponent, are infinite.  So, for example, if the tail of the mutual fund size distribution follows Zipf's law as hypothesized by Gabaix et al., i.e. if it were a power law with tail exponent one, this would imply nonexistence of the mean.  In this case the sample estimator fails to converge because  the tails are so heavy that with significant probability a single fund can be larger than the rest of the sample combined.  This is true even in the limit as the sample size goes to infinity.  Thus power law tails imply a very high degree of concentration.

Instead, empirical analysis shows that the tail of the mutual fund size distribution is not a power law, and is  well-approximated by a lognormal \citep{Schwarzkopf10a}.  Thus, while the distribution is heavy tailed, it is not as heavy tailed as it would be if the distribution were a power law.   The key difference is that for a log-normal all of the moments exist.

This naturally leads to the question of what economic factors determine the tail properties of the mutual fund distribution.  There are two basic types of explanation.  One type of explanation is based on a detailed description of investor choice, and another is based on efficient markets, which predicts that growth should be random, and that the causes can be understood in terms of a simple random process description of entry, exit and growth.   Of course market efficiency depends on investor choice, but the key distinction is that the random process approach does not depend on any of the details, but rather only requires that no one can make superior investments based on simple criteria, such as size.

Explanations based on investor choice can in turn be divided into two types: Rational and behavioral.  For example, Berk and Green [\citeyear{Berk04}] have proposed that investors are rational, making investments based on past performance.  Their theory implies that the distribution of fund size is determined by the skill of mutual fund managers and the dependence of transaction costs on size.  If we assume, for example, that the transaction cost is a power law (which includes linearity) if the distribution of fund size is log-normal, then it is possible to show that the distribution of mutual fund skill must also be log-normal.  Unfortunately, without a method of measuring skill this is difficult to test.

Another type of explanation is behavioral, i.e. that investors are strongly influenced by factors such as advertising, fees, and investment fads\footnote{
\citet{Barber05} have found that investors flows are correlated to marketing and advertising while they are not correlated to the expense ratio.}. 
We strongly suspect that this is true, and that they play an important role in determining the size of individual funds.  The question we investigate here is not whether such effects exist, but whether they are essential to explain the form of the distribution.

The alternative is that the details of investor choice don't matter, and that the distribution of fund size is driven by market efficiency, which dictates an approach based on the random process of entry, exit and growth.  The random process approach was originally pioneered as an explanation for firm size by Gibrat, Simon and Mandelbrot, and is popular in the industrial organization literature\footnote{
 For past stochastic models  see \citep{gibrat-1931,simon-1955,simon-1958,mandelbrot-1963,ijiry-1977,sutton-1997,gabaix-2003-mit,gabaix-2003-nature}.}.
The basic idea is that while details of investor choice 
are surely important in determining the size of {\it individual} funds, the details may average out or be treatable as noise, so that in aggregate they do not matter in shaping the overall size distribution.
 
On the face of it, however, there seems to be a serious problem with this approach.  Under simple assumptions about the entry, exit and growth of fund size,  \cite{gabaix-2003-mit} showed that the steady state solution is a power law;  a similar argument is described in \cite{Montroll82} and \cite{Reed01}\footnote{
For a review on similar generative models see  \cite{Mitzenmacher03}.}. 
  As already mentioned, however, the upper tail of the empirical distribution is a log-normal, not a power law.  Thus there would seem to be a contradiction.  Apparently either the correct random process is more complicated, or this whole line of attack fails.

We show here that the central problem comes from considering only the steady state (i.e. infinite time) solution.  We study the same equations considered by Gabaix et al. and Reed, but we find a more general time-dependent solution, and show that the time required to reach steady state is very long.  The mutual fund industry is rapidly growing and, even if the growth process had been stationary over the last few decades, not enough time has elapsed to reach the stationary solution for the fund size distribution.  In the meantime the solution is well approximated by a log-normal.  This qualitative conclusion is very robust under variations of the assumptions.  In contrast to the hypothesis of Berk and Green, it does not depend on details such as the distribution of investor skill -- the log-normal property emerges automatically from market efficiency and the random multiplicative nature of fund growth.

To test our conjectures more quantitatively we study the empirical properties of entry, exit and growth of mutual funds, propose a more accurate model than those previously studied, and show it makes a good prediction of the empirically observed fund size distribution.   The model differs from previous models in that it incorporates the fact that the relative growth rate of funds slows down as they get bigger\footnote{
For work on the size dependence of firm growth rate fluctuations see\citep{stanley-1995,stanley-1996,Amaral97,Bottazzi01,Bottazzi03a,Bottazzi05,dosi-2005,Defabritiis03}.}.
This makes the time needed to approach the steady state solution even longer:  Whereas the relaxation time for the size-independent diffusion model is several decades, for the more accurate size-dependent model it is more than a century.

Market efficiency is the key economic principle that makes the random process model work, and dictates many of its properties.  It enters the story in several ways.  (1) The fact that stock market returns are essentially random implies that growth fluctuations are random, for two reasons:  (a) Without inflows and outflows, under the principle that past returns are not indicative of future returns, fund growth is random.  (b) Although investors chase past returns, since what they are chasing is random, fund growth due to inflow and outflow is random on sufficiently long time scales.  (2) Efficiency dictates that mutual fund performance must be independent of size.  Thus as mutual funds randomly diffuse through the size space, there is no pressure pushing them toward a particular size.  (3) Efficiency, together with the empirical fact that the relative importance of fund inflows and outflows diminishes as funds get bigger, implies that the mean growth rate and the growth diffusion approach a constant in the large size limit.  As we show, this shapes the long-term properties of the size distribution.  All of these points are explained in more detail in Section~\ref{sec:size_change}.

Market efficiency makes mutual funds unusual relative to most other types of firms.  For most firms, in the large size limit the mean and standard deviation of the growth rate are empirically observed to decay to zero.  For mutual funds, in contrast, due to market efficiency they both approach a positive limit.  This potentially affects the long-term behavior:  Most firms approach a solution that is thinner than a log-normal, i.e under stationary growth conditions  their tails are getting thinner with time, whereas mutual funds approach a power law, so their tails are getting fatter with time.  Nonetheless, as we have already mentioned, even under stationary growth conditions the approach to steady-state takes so long that this is a moot point.

At a broader level our work here shows how the non-stationarity of market conditions can prevent convergence to an ``equilibrium" solution.   Nonetheless, even under stationary conditions the random process model usefully describes the time-dependent relationships between entry, exit and growth phenomena on one hand and size on the other hand.  While we cannot show that the random process model is the only possible explanation, we do show that it provides a good explanation\footnote{
While variations in the assumptions about the random process preserve certain qualitative conclusions, such as the log-normal character of the upper tail, we found that getting a good fit to the data requires a reasonable degree of fidelity in the modeling process.  The size-dependent nature of the diffusion process, for example, is quite important.}.
The conditions for this are robust, depending only on market efficiency, without the stronger requirements of perfect rationality, or the complications of mapping out the idiosyncrasies of human behavior.   

 

The paper is organized as follows. In Section~\ref{section_model} we develop the standard exit and entry model. Section~\ref{section_N} presents the time-dependent solution for the number of funds, Section ~\ref{section_solution} presents the time-dependent solution for the size distribution, and Section~2.3 introduces a size-dependent model.
Section~\ref{data_set} describes our data. In Section~\ref{section_empiric_justification} we perform an empirical analysis to justify our assumptions and to calibrate the model.
In Section ~\ref{sec:comparison}  we present simulation results of the proposed model and compare them to the empirical data. 
Finally Section~\ref{conclusions} presents our conclusions.


\section{Model}\label{section_model}

Our central thesis in this paper is that due to market efficiency the mutual fund size distribution can be explained by a stochastic process governed by three key underlying processes: the size change of existing mutual funds, the entry of new funds and the exit of existing funds.  In this section we introduce the standard diffusion model and derive a time-dependent solution for the special case when the diffusion process has constant mean and variance.  We then make a proposal for how to model the more general case where the mean and variance depend on size.

The aim of the model we develop here is to describe the time evolution of the size distribution, that is, to solve for the probability density function $p(\w,t)$ of funds with  size $s$ at time $t$, where $\w = \log s$.  The size distribution can be written as
\begin{equation}
p(\w,t)=\frac{n(\w,t)}{N(t)},
\end{equation}
where $n(\w,t)$ is the number of funds at time with logarithmic size $\w$ and $N(t)=\int n(\w,t)\md \w$ is the total number of funds at time $t$.
To simplify the analysis we solve separately for the total number of funds $N(t)$ and for the number
density $n(\w,t)$. 

\subsection{Dynamics of the total number of funds}\label{section_N}

As we will argue in Section~\ref{section_empiric_justification}, the total number of funds as a function of time can be modeled as
\begin{equation}
\frac{dN}{dt} = \nu - \lambda N
\end{equation}
where $\nu$ is the rate of creating new funds and $\lambda$ is the exit rate of existing funds.  Under the assumption that $\nu$ and $\lambda$ are constants this has the solution
\begin{equation}\label{eq_Nt}
N(t)=\frac{\nu}{\lambda}\left(1-\e^{-\lambda t}\right)\theta(t),
\end{equation}   
where $\theta(t)$ is a unit step function at $t = 0$, the year in which the first funds enter. This solution has the surprising property that the dynamics only depend on the fund exit rate $\lambda$, with a characteristic timescale $1/\lambda$.  For example, for $\lambda \approx 0.09$, as estimated in Section~\ref{section_empiric_justification}, the timescale for $N(t)$ to reach its steady state is only roughly a decade.  An examination of Table~\ref{table} makes it clear, however, that $\nu = \mbox{constant}$ is not a very good approximation.  Nonetheless, if we crudely use the mean creation rate $\nu \approx 900$ from Table~\ref{table} and the fund exit rate $\lambda \approx 0.09$ estimated in Section~\ref{section_empiric_justification}, the steady state number of funds should be about $N \approx 10,000$, compared to the $8,845$ funds that actually existed in 2005.  Thus this gives an estimate with the right order of magnitude.

The important point to stress is that the dynamics for $N(t)$ operate on a different timescale than that of $n(\omega, t)$.  As we will show in the next section the characteristic timescale for $n(\omega, t)$ is much longer than that for $N(t)$.

\subsection{Solution for the number density $n(\w,t)$}\label{section_solution}

We define and solve the time evolution equation for the number density $n(\w,t)$.  The empirical justification for the hypotheses of the model will be given in Section~\ref{section_empiric_justification}.
The hypotheses are: 

\begin{itemize}
\item
The entry process is a Poisson process with rate $\nu$, such that at time $t$ a new fund enters the industry with a probability $\nu\mathrm{d}t$ and (log) size $\w$ drawn from a distribution $f(\w,t)$. We approximate the entry size distribution as a log-normal distribution in the fund size $s$, that is a normal distribution in $\w$ given by 
\begin{equation}\label{entry_distribution}
f(\w,t)=\frac{1}{\sqrt{\pi\sigma_{\w}^2}}\exp\left(-\frac{(\w-\w_0)^2}{\sigma_{\w}^2}\right)\theta(t-t_0),
\end{equation}
where $\w_0$ is the mean log size of new funds and $\sigma_\w^2$ is its variance. $\theta(t-t_0)$ is a unit step function ensuring no funds funds enter the industry before the initial time $t_0$.
\item
The exit process is a Poisson process such that at any time time $t$ a fund exits the industry with a size independent probability $\lambda\md t$. 
\item
The size change is approximated as a (log) Brownian motion with a size dependent drift and diffusion term
\begin{equation}\label{geometric_random_walk}
\mathrm{d}\w=\mu(\w)\mathrm{d}t+\sigma(\w)\mathrm{d}W,
\end{equation}
where $\mathrm{d}W$ is an  i.i.d random variable drawn from a zero mean and unit variance normal distribution.
\end{itemize}

Under these assumptions the forward Kolmogorov equation (also known as the Fokker-Plank equation) defining the time evolution of the number density \citep{Gardiner04} is given by  
 \begin{equation}\label{fokker-plank}
 \frac{\partial }{\partial t} n(\w,t)= \nu f(\w,t) -\lambda n(\w,t) - \frac{\partial}{\partial \w}[\mu(\w)n(\w,t)]%
+\frac{\partial^2}{\partial \w^2}[D(\w) n(\w,t)],
\end{equation}
where $D(\w)=\sigma(\w)^2/2$ is the size diffusion coefficient.
The first term on the right describes the entry process, the second describes the fund exit process and the third and fourth terms describe the change in size of a existing funds.

\subsubsection{Approximate solution for large funds}

To finish the model it is necessary to specify the functions $\mu(\w)$ and $D(\w)$.   It is convenient to define the relative change in a fund's size $\Delta_s(t)$ as
\begin{equation}\label{ds}
 \Delta_s(t) =\frac{s(t+1)-s(t)}{s(t)},
 \end{equation}
 such that drift and diffusion parameters in our model are given by 
 \[\mu(\w)=\mathrm{E}[\log(1+\Delta_s)] \,\,\,\,\,\,\,\,\,\, D(\w)=\frac{1}{2} \mathrm{Var}[\log(1+\Delta_s)].\] 
 The relative change can be decomposed into two parts: the return $\Delta_r$ and the fractional investor money flux $\Delta_f (t)$, which are simply related as
\begin{equation}\label{ds_decomposed}
 \Delta_s(t)=\Delta_f(t)+\Delta_r(t).
 \end{equation}
 The return $\Delta_r$ represents the return of the fund to its investors, defined as
\begin{equation}\label{dr}
\Delta_r(t)=\frac{NAV(t+1)-NAV(t)}{NAV(t)},
\end{equation}
where $NAV(t)$ is the Net Asset Value at time $t$.  The fractional money flux $\Delta_f (t)$ is the change in the fund size by investor deposits or withdrawals, defined as
\begin{equation}\label{df}
\Delta_f(t)=\frac{s(t+1)-[1+\Delta_r(t)]s(t)}{s(t)}.
\end{equation}

In Section~\ref{section_empiric_justification} we will demonstrate empirically that the returns $\Delta_r$ are independent of size, as they must be for market efficiency.   In contrast the money flux $\Delta_f$ decreases monotonically with size.  In the large size limit the returns $\Delta_r$ dominate, and thus it is reasonable to treat $\mu(s)$ as a constant, $\mu= \mu_\infty$.   Market efficiency also implies that in the large size limit the standard deviation $\sigma(s)$ is a constant, i.e. $\sigma= \sigma_\infty$.  Otherwise investors would be able to improve their risk adjusted returns by simply investing in larger funds. 

With these approximations the evolution equation becomes
\begin{equation}
 \frac{\partial }{\partial t} n(\w,t)= \nu f(\w,t) -\lambda n(\w,t) -\mu\frac{\partial}{\partial \w}n(\w,t)%
+D\frac{\partial^2}{\partial \w^2}n(\w,t),
\label{fokker-plank_large_size}
\end{equation}
In this and subsequent equations, to keep things simple we use the notation $D=\sigma_\infty^2/2$ and $\mu = \mu_\infty$.

The exit process is particularly important, since it is responsible for thickening the upper tail of the distribution. The intuition is as follows:  Since each fund exits the industry with the same probability, and since there are more small funds than large funds, more small funds exit the industry. This results in relatively more large funds, making the distribution heavy-tailed.   As we will now show this results in the distribution evolving from a log-normal upper tail to a power law upper tail. 
In contrast, the entry process is not important for determining the shape of the distribution, and influences only the total number of funds $N$. This is true as long as the entry size distribution $f(\w,t)$ is not heavier-tailed than a lognormal, which is supported by the empirical data.

In the large size limit the solution for an arbitrary entry size distribution $f$ is given by 
\begin{equation}
n(\w,t)=\nu \int_{-\infty}^{\infty}\int_0^{t}\exp^{-\lambda t'}\frac{1}{\sqrt{4 \pi D t'}}\exp\left[-\frac{(\w-\w' -\mu t')^2}{4 D t'}\right] f(\w',t-t')\, \md t'\md \w'.
\end{equation} 
Stated in words,  a fund of size $\w'$ enters at time $t-\tau$ with probability $f(\w',t-\tau)$. The fund will survive to time $t$ with a probability $\exp(-\lambda \tau)$ and will have a size $\w$ at time $t$ with a probability according to  (\ref{solution_no_exit_or_entry}).

If funds enter the industry with a constant rate $\nu$ beginning at $t=0$, with a log-normal entry size distribution $f(\w,t)$ centered around $\w_0$ with width $\sigma_\w$  as given by (\ref{entry_distribution}), the size density can be shown to be
\begin{eqnarray}
n(\w,t)&=&\frac{\nu\mu}{4 \sqrt{\gamma } D} \exp\left[(\gamma +\frac{1}{4})\frac{ \sigma_\w^2}{2}-\sqrt{\gamma } \left|\frac{\sigma_\w^2}{2}+ \frac{\mu}{D}\left(\w-\w_0\right)\right|+\frac{\mu}{2D} \left(\w-\w_0\right)\right] \nonumber\\%
  &\,&\times\left( A +\exp\left[\sqrt{\gamma }|\sigma_\w^2+2 \frac{\mu}{D} \left(\w-\w_0\right)|\right] B\right).
\end{eqnarray}
The parameters A, B and $\gamma$ are defined as
\begin{equation}\label{eq_gamma}
\gamma=\sqrt{\frac{1}{4}+\frac{\lambda D}{\mu^2}},
\end{equation}
\begin{eqnarray}
A&=&\mathrm{Erf}\left[\frac{\left|\frac{\sigma_\w^2}{2}+ \frac{\mu}{D}\left(\w-\w_0\right)\right|- \sqrt{\gamma } \sigma_\w^2}{ \sqrt{2} \sigma_\w}\right]\\%
&& -\mathrm{Erf}\left[\frac{\left|\frac{\sigma_\w^2}{2}+ \frac{\mu}{D}\left(\w-\w_0\right)\right|- \sqrt{\gamma } \left(\sigma_\w^2+2\frac{\mu^2}{D}t\right)}{\sqrt{2} \sqrt{\sigma_\w^2+2\frac{\mu^2}{D} t}}\right]\nonumber
\end{eqnarray}
and
\begin{eqnarray}
B&=& \mathrm{Erf}\left[\frac{ \sqrt{\gamma } \left(\frac{\sigma_\w^2}{2}+\frac{\mu^2}{D}t\right)+|\frac{\sigma_\w^2}{2}+ \frac{\mu}{D}\left(\w-\w_0\right)|}{\sqrt{2} \sqrt{\sigma_\w^2+2 \frac{\mu^2}{D}t}}\right]\\
&&-\mathrm{Erf}\left[\frac{ \sqrt{\gamma } \sigma_\w^2+\left|\frac{\sigma_\w^2}{2}+ \frac{\mu}{D}\left(\w-\w_0\right)\right|}{ \sqrt{2} \sigma_\w}\right],\nonumber
\end{eqnarray}
where $\mathrm{Erf}$ is the error function, i.e. the integral of the normal distribution.

Approximating the distribution of entering funds as having zero width simplifies the solution.  Let us define a large fund as one with $\w\gg\w_0$, where $\w_0$ is the logarithm of the typical entry size of one million USD.  For large funds we can approximate the lognormal distribution as having zero width, i.e. all new funds have the same size $\w_0$.   The number density is then given by 
\begin{eqnarray}\label{time_dependent_solution}
n(\w,t)&=&\frac{ \nu D}{4 \sqrt{\gamma }\mu ^2}e^{\frac{1}{2} \frac{\mu}{D}\left(\w -\w_0\right)}%
\Bigg[e^{-\sqrt{\gamma } \frac{\mu}{D}|\w -\w_0|}\left(1+ \mathrm{erf}\left[\sqrt{\frac{\gamma\mu^2 t}{D}}-\frac{|\w -\w_0|}{2 \sqrt{D t}} \right]\right)\nonumber\\%
&&- e^{\sqrt{\gamma }\frac{\mu}{D} |\w -\w_0|}%
 \left(1-\mathrm{erf}\left[\mu\sqrt{\frac{t}{D}}(\frac{1}{2}+\sqrt{\gamma})\right]\right)\Bigg].
 \label{solution_constant_size}
\end{eqnarray}
Since $\gamma > 1/4$ (\ref{eq_gamma}), the density vanishes for both $\w\to \infty$ and $\w\to -\infty$.

\subsubsection{Steady state solution for large funds}

The steady state solution for large times is achieved by taking the $t\to\infty$  limit of (\ref{solution_constant_size}), which gives
\begin{equation}\label{n_delta_s}
n(\w)=\frac{\nu}{2\mu\sqrt{\gamma}}
 \exp\frac{\mu}{D}\left(\frac{\w-\w_0}{2}-\sqrt{\gamma}|\w-\w_0|\right).
\end{equation}
Since the log size density (\ref{n_delta_s}) has an exponential upper tail \mbox{$p(\w)\sim \exp(-\zeta_s \w)$} and \mbox{$s=\exp(\w)$} the CDF for $s$ has  a power law tail with an exponent\footnote{
To calculate the tail exponent of the density correctly one must change variables through $p(s)=p(\w)\frac{\mathrm{d}\w}{\mathrm{d}s}\sim s^{-\zeta_s-1}$. This results in a CDF with a tail exponent of $\zeta_s$.}
 $\zeta_s$, i.e.
\begin{equation}
P(s>X)\sim X^{-\zeta_s}.
\end{equation}
Substituting for the parameter $\gamma$ using Eq.~(\ref{eq_gamma}) for the upper tail exponent yields 
\begin{equation}\label{eq_zeta}
\zeta_s=\frac{-\mu+\sqrt{\mu^2+4D\lambda}}{2D}.
\end{equation}
Note that this does not depend on the creation rate $\nu$.  Using  the average parameter values in Table~\ref{table_fit} the asymptotic exponent has the value
\begin{equation}
\zeta_s=1.2\pm0.6.
\end{equation}
This suggests that if the distribution reaches steady state it will follow Zipf's law, which is just the statement that it will be a power law with $\zeta_s\approx1$.  As discussed in the introduction, this creates a puzzle, as the empirical distribution is clearly log-normal \citep{Schwarzkopf10a}.    

\subsubsection{Timescale to reach steady state}

Since we have a time dependent solution we can easily estimate of the timescale to reach steady state.   The time dependence in Eq.~\ref{time_dependent_solution} is contained in the arguments of the error function terms on the right.   When these arguments become large, say larger than 3, the solution is roughly time independent, and can be written as 
\begin{equation}
t > \frac{9 D}{4 \gamma \mu ^2}\left(1+\sqrt{1+\frac{2}{9} \frac{ \sqrt{\gamma\mu^2} }{D} \left|\w-\w_0\right| }\right)^2.
\label{timeScale1}
\end{equation}
Using the average values in Table~\ref{table_fit} in units of months $\mu=\mu_\infty \approx 0.005$, $D=\sigma_\infty^2/2$ and $\sigma_{\infty} \approx 0.05$.  This gives
\[
t >180 \left(1+\sqrt{1+0.7\left|\w-\w_0\right|  }\right)^2,
\]
where the time is in months.  Plugging in some numbers from Table~\ref{table_fit} makes it clear that the time scale to reach steady state is very long. For instance, for funds of a billion dollars it will take about 170 years for their distribution to come within 1 percent of its steady state.
This agrees with the empirical observation that there seems to be no significant fattening of the tail in the fifteen years from 1991 - 2005.  Note that the time required for the distribution $n(\w, t)$ to reach steady state for large values of $\w$ is much greater than that for the total number of funds $N(t)$ to become constant.

During the transient phase the solution remains approximately log-normal for a long time.  If funds only change in size and no funds enter or exit, then the resulting distribution is normal
\begin{equation}\label{solution_no_exit_or_entry}
\tilde{n}(w,t)=\frac{1}{\sqrt{4 \pi D t}}\exp\left[-\frac{(\w-\mu t)^2}{4 D t}\right],
\end{equation}
which corresponds to a size distribution $p(s)$ with a lognormal upper tail.  While the exit process acts quickly in changing the total number of funds, it acts slowly in changing the shape. This is the key reason why the distribution remains approximately log-normal for so long.
 
\subsection{A better model of size dependence \label{muAndSigma}}

The mean rate of growth and diffusion are in general size dependent.   We hypothesize that the mean growth rate $\mu(s)$ and the standard deviation $\sigma(s)$ are the sum of a power law and a constant, of the form
\begin{eqnarray}\label{eq_mu_sigma_power_law}
\sigma_s(s)&=&\sigma_0s^{-\beta} +\sigma_{\infty} \\
\mu_s(s)&=&\mu_0s^{-\alpha}+\mu_{\infty}.\nonumber
\end{eqnarray}
The constant terms come from mutual fund returns (neglecting inflow or outflow of funds), and must be constant due to market efficiency, as explained in more detail in Section~\ref{sec:size_change}.  The power law terms, in contrast, are due to the flow of funds in and out of the market.  There is a substantial literature of proposed theories for this, including ours\footnote{
There has been a significant body of work attempting to explain the heavy tails in the growth rate of firms and the associated size dependence in the diffusion rate.  See \citep{Amaral97,Buldyrev97,Amaral98,Defabritiis03,Matia04,Bottazzi01,Sutton01,Wyart03,Bottazzi03b,Bottazzi05,Fu05,Riccaboni08,Podobnik08}.   Our theory argues for an additive replication model, and produces predictions that fit the data extremely well for a diverse set of different phenomena, including mutual funds \citep{Schwarzkopf10b}.  We argue that the fundamental reason for the power tails is the influence network of investors.}.
We present the empirical evidence for the power law hypothesis and explain the role of efficiency in more detail in Section~\ref{sizeDependence}.

The functional form given above for the size dependence can be used to make a more accurate diffusion model.  The non vanishing drift $\mu_\infty>0$ and diffusion terms $\sigma_\infty>0$ are essential for the distribution to evolve towards a power law.  As already mentioned, due to market efficiency $E[\Delta_r(s)]$ must be independent of $s$, and since $E[\Delta_f(s)]$ is a decreasing function of $s$, for large $s$  $ \mu(s) = E[\Delta_r(s)] + E[\Delta_f(s)] = \mu_\infty > 0$.  This distinguishes mutual funds from other types of firms, which are typically observed empirically to have $\mu_\infty = \sigma_\infty = 0$ \citep{stanley-1996,Matia04}.   Assuming that other types of firms obey similar diffusion equations to those used here, it can be shown that the resulting distribution has a stretched exponential upper tail, which is much thinner than a power law\footnote{
A stretched exponential is of the form $p(x) \sim \exp(a x^{-b})$, where $a$ and $b$ are positive constants. There is some evidence in the empirical data that the death rate $\lambda$ also decays with size.  However, in our simulations we found that this makes very little difference for the size distribution as long as it decays slower than the distribution of entering funds, and so in the interest of keeping the model parsimonious we have not included this effect.}.
%
 
\section{Data Set}\label{data_set}

We test our model against the  CRSP Survivor-Bias-Free US Mutual Fund Database. Because we have daily data for each mutual fund, this database enables us to investigate the mechanism of fund entry, exit and growth to calibrate and test our model\footnote{
Note that we treat mergers as the dissolution of both original firms followed by the creation of a new (generally larger) firm.  This increases the size of entering firms but does not make a significant difference in our conclusions.}.
We study the data from 1991 to 2005\footnote{
There is data on mutual funds starting in 1961, but prior to 1991 there are very few entries.  There is a sharp increase in 1991, suggesting incomplete data collection prior to 1991.  }.
We define an equity fund as one whose portfolio consists of at least $80\%$ stocks.  The 
results are not qualitatively sensitive to this, e.g. we get essentially the same results even if we use all funds.    
The data set has monthly values for the Total Assets Managed (TASM) by the fund and the Net Asset Value (NAV).  We define the size $s$ of a fund to be the value of the TASM, measured in millions of US dollars and corrected for inflation relative to July 2007.  Inflation adjustments are based on the Consumer Price Index, published by the BLS.
In Table~\ref{table} we provide summary statistics of the data set and as seen there the  total number of equity funds increases roughly linearly in time, and the number of funds in the upper tail $N_{tail}$ also increases.
%
 
\begin{sidewaystable}
\begin{center}
\footnotesize
\begin{tabular}{c|c|c|c|c|c|c|c|c|c|c|c|c|c|c|c||c|c}
\hline
variable & 91&92&93&94&95&96&97&98&99&00&01&02&03&04&05&mean&std \\
\hline \hline
$N$ & 372 & 1069 & 1509 & 2194 & 2699  &  3300 & 4253 & 4885 & 5363 & 5914 & 6607 & 7102
&7794& 8457 & 8845& - & -  \\
\hline\hline
E$[s]$ (mn)&810&  385  &  480   & 398   & 448  &  527  &  559  &  619  &  748  &  635  &  481  &  335  & 425  &  458 &  474 & 519 & 134\\
 \hline 
 Std$[s]$ (bn) &1.98   & 0.99  & 1.7  &  1.66 & 1.68  &  2.41  &  2.82  &  3.38   & 4.05  &  3.37 &  2.69  &  1.87  & 2.45  &  2.64   & 2.65 & 2.42& 0.8\\
\hline \hline
E$[\w]$&5.58  &  4.40  &  4.40   & 3.86   & 3.86  &  3.91  &  3.84  &  3.85  &  4.06  &  3.97  &  3.60  &  3.37  & 3.55  &  3.51  &  3.59&3.96&0.54\\
 \hline 
 Std$[\w]$ &1.51   & 1.98  &  2.09  &  2.43 & 2.50  &  2.46  &  2.50  &  2.51   & 2.46  &  2.45 &  2.63  &  2.42  & 2.49  &  2.59   & 2.50 & 2.34& 0.29\\
\hline \hline
$\mu (10^{-3})$ & 26 & 42 & 81 & 46 & 72 & 67 & 58 & 39 & 39 & 20 & -3 & -10 & 50 & 18 & 30.5 & 38.5& 26 \\
\hline
$\sigma (10^{-1})$& 0.78& 2.0 & 2.6 & 3.0 & 2.3 & 2.8 & 2.9 & 3.0 & 2.8 & 2.8 & 2.5 & 2.4 & 2.4 & 2.4 & 2.5 & 2.5 & 0.55 \\
\hline\hline
$N_{\tt exit}$ &0 &41& 45& 61&139 &115&169&269&308&482&427 &660&703&675&626& - & - \\
\hline 
$N_{\tt enter}$&185&338&581&783&759&885&1216&1342&1182&1363&1088&1063&1056&796&732&891 & 346 \\
\hline 
\end{tabular}
\normalsize 
\end{center}
\label{default}
\caption{\label{table}
Summary statistics and parameter values for equity funds defined such that the portfolio contains a fraction of at least $80\%$ stocks. 
The values for each of the parameters (rows) were calculated for each year (columns). The mean and standard deviation are evaluated across the different years.  \newline
$N$ - the number of equity funds existing at the end of each year. \newline
$E[\w]$ - the mean log size of funds existing at the end of each year. \newline
$Std[\w]$ - the standard deviation of log sizes for funds existing at the end of each year. \newline
$E[s]$ - the mean size (in millions) of funds existing at the end of each year. \newline
$Std[s]$ - the standard deviation of sizes (in billions) for funds existing at the end of each year. \newline
$\mu$ - the drift term for the geometric random walk (\ref{geometric_random_walk}), computed for monthly changes. \newline
 $\sigma$ - the standard deviation of the mean zero Wiener process (\ref{geometric_random_walk}), computed for monthly changes. \newline
$N_{\tt exit}$ - the number of equity funds exiting the industry each year. \newline
 $N_{\tt enter}$ - the number of new equity funds entering the industry in each year.   }
\end{sidewaystable}

\section{Empirical investigation of size dynamics}\label{section_empiric_justification}

In this section we empirically investigate the processes of entry, exit and growth, providing empirical justification and calibration of the model described in Section~2.

 \subsection{Fund entry}\label{sec:entry}
 
We begin by examining the entry of new funds. We investigate both the number $N_{\tt enter}(t)$ and size $s$ of funds entering each year.  We perform a linear regression of $N_{\tt enter}(t)$ against the number of existing funds $N(t-1)$, yielding slope $\alpha=0.04 \pm 0.05$ and intercept $\beta =750 \pm 300$.   The slope is not statistically significant, justifying the approximation of entry as a Poisson process with a constant rate $\nu$, independent of $N(t)$.

\begin{figure}
\begin{center}
\includegraphics[width=10cm]{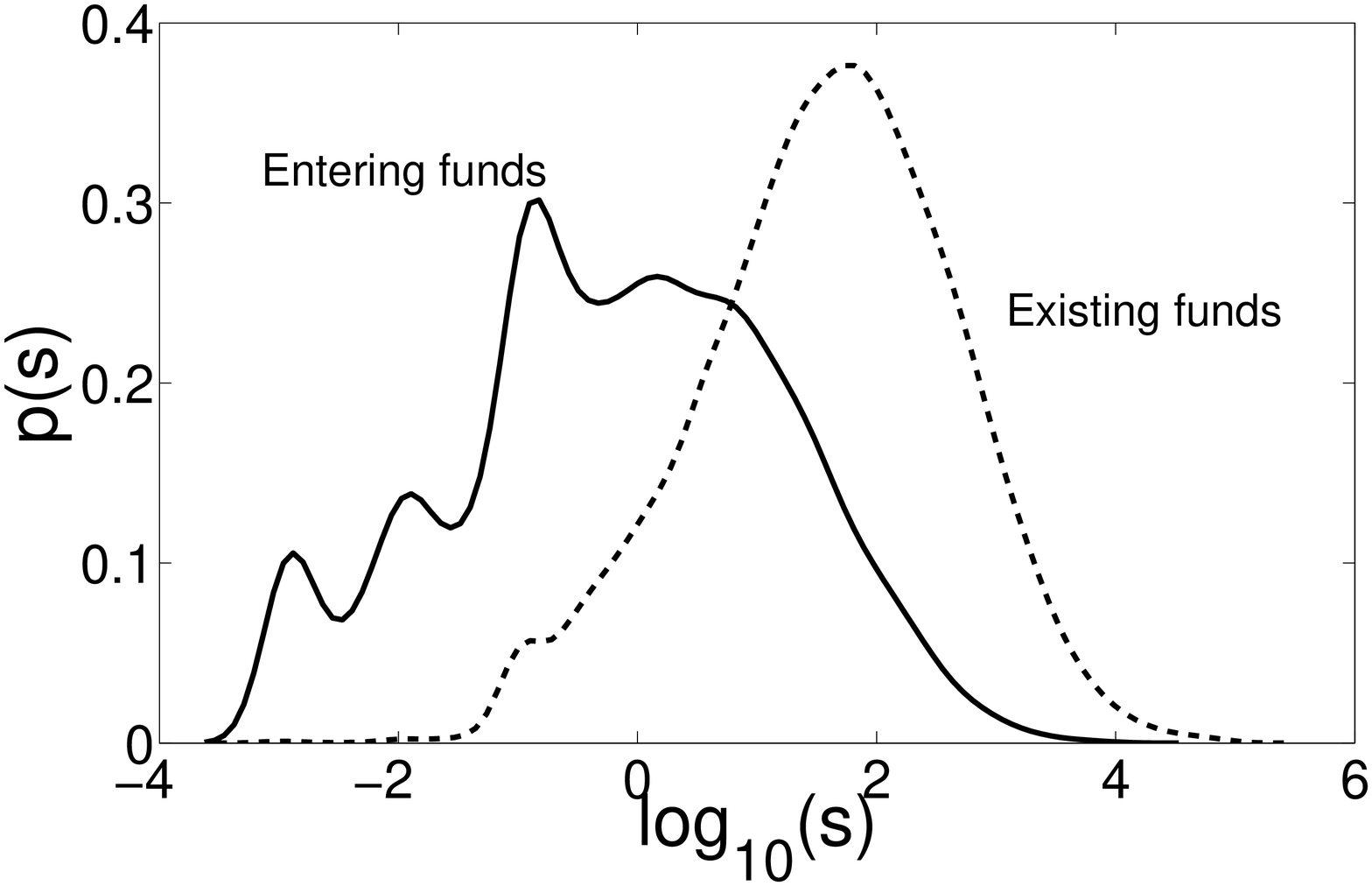}
\caption{\label{Winit}
The probability density for the size $s$ of entering funds in millions of dollars (solid line) compared to that of all funds (dashed line) including all data for the years 1991 to 2005.  The densities were estimated using a gaussian kernel smoothing technique.}
\end{center}
\end{figure}
The size of entering funds is more complicated.  
In Figure~\ref{Winit} we compare the distribution of the size of entering funds $f(s)$ to that of all existing funds.  The distribution is somewhat irregular, with peaks at round figures such as ten thousand, a hundred thousand, and a million dollars. The average size\footnote{
When discussing the average size one must account for the difference between the average log size and the average size: Due to the heavy tails the difference is striking. The average entry log size E$[\w_c]\approx 0$, corresponding to a fund of size one million, while if we average over the entry sizes E$[s_c]=\mathrm{E}[\e^{\w_c}]$, we get an average entry size of approximately 30 million. For comparison, both the average size and the average log size of existing funds are quoted in Table~\ref{table}. }
 of entering funds is almost three orders of magnitude smaller than that of existing funds, making it clear that the typical surviving fund grows significantly after it enters.  It is clear that the distribution of entering funds is not important in determining the upper tails\footnote{
 In Section~\ref{section_solution} we showed that the entry process is not important as long as the tails of the entry distribution $f$ are sufficiently thin.  We compared the empirical $f$ to a log-normal and found that the tails are substantially thinner.}.
The value of the mean log size and its variance are calculated from the data for each period as summarized in Table~\ref{table_fit}.  

Thus the empirical data justifies the approximation of entry as a Poisson process in which an average of $\nu$ funds enter per month, with the size of each fund drawn from a distribution $f(\w,t)$.

 \subsection{Fund exit}

\begin{figure}
\begin{center}
\includegraphics[width=10cm]{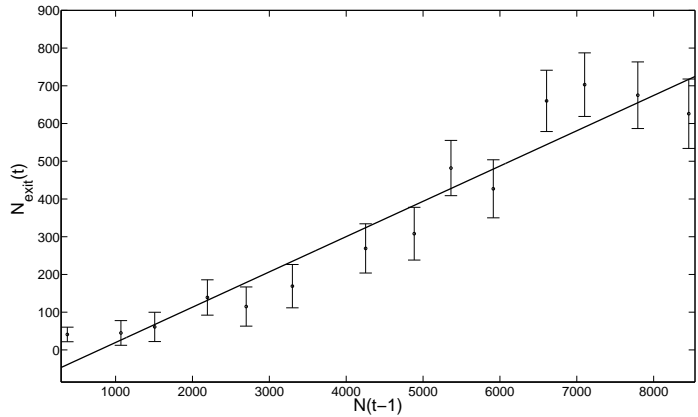}
\caption{\label{Nfinal_vs_N}
The number of equity funds  exiting the industry  $N_{\tt exit}(t)$ in the year $t$ as a function of the total number of funds existing in the previous year, $N(t-1)$. The plot is compared to a linear regression (full line). 
The error bars are calculated for each bin under a Poisson process assumption, and correspond to the square root of the average number of funds exiting the industry in that year. 
}
\end{center}
\end{figure}

Unlike entry, fund exit is of critical importance in determining the long-run properties of the fund size distribution.  In Figure~\ref{Nfinal_vs_N} 
we plot the number of exiting funds $N_{\tt exit}(t)$ as a function of the total number of  funds existing in the previous year, $N(t-1)$.   There is a good fit to a line of slope $\lambda$, which on an annual time scale is $\lambda=0.092 \pm 0.030$.  
This justifies our assumption that fund exit is a Poisson process with constant rate $\lambda$.
\subsection{Fund growth}\label{sec:size_change}
We first test the i.i.d and normality assumptions of the diffusion growth model, and then test to demonstrate the size dependence of the growth process that we proposed in Section~\ref{muAndSigma}.  We also discuss the diverse roles that efficiency plays in shaping the random process for firm growth in more detail.

\subsubsection{Justification for the diffusion model\label{justification}}

In the absence of entry or exit we have approximated the growth of existing funds as a multiplicative Gibrat-like process\footnote{
A Gibrat-like process is a multiplicative process in which the size of the fund at any given time is given as a multiplicative factor times the size of the fund at a previous time. In Gibrat's law of proportionate effect  \citep{gibrat-1931} the multiplicative term depends  linearly on size while here we allow it to have any size dependence.}
satisfying a random walk in the log size $\w$.   This implicitly assumes that $\Delta_s$ is an i.i.d normal random variable.

The assumption of independence is justified by market efficiency, which requires that the returns $\Delta_r$ of a given fund should be random \citep{bollen05,carhart97}.  Under the decomposition of the total growth as $\Delta_s = \Delta_r + \Delta_f$, as demonstrated in the next sub-section, in the large size limit the returns $\Delta_r$ dominate, so under market efficiency the i.i.d. assumption is automatically valid.

This is not so obvious for smaller size firms, where the money flux $\Delta_f$ dominates the total growth $\Delta_s$.  It is well known that investors chase past performance\footnote{
For empirical evidence that investors react to past performance see
\citep{remolona-1997,busse-2001,chevalier-1997,sirri-1998,delguercio-2002,bollen-2007}. }.
Even though the past performance they are chasing is random, if they track a sufficiently long history of past returns, this can induce correlations.
This causes correlations in the money flux $\Delta_f$, which in turn induces correlations in the total size change $\Delta_s$.  

To test whether such correlations are strong enough to cause problems with the random process hypothesis, we perform cross-sectional regressions of the form
\begin{equation}
\Delta_f(t)=\beta+\beta_1\Delta_r(t-1)+\beta_2\Delta_r(t-2)+\ldots +\beta_6\Delta_r(t-6) + \xi(t),
\label{performanceRegression}
\end{equation}
where $\xi(t)$ is a noise term.  The results are extremely noisy; for example, when we perform separate regressions in five different periods, eight of the thirty possible coefficients $\beta_i$ shown in Table~\ref{corr_table} are negative and only two of them are significant at the two standard deviation level.  We also perform direct tests of the correlations in $\Delta_f$ and we find that they are small.  This justifies our use of the i.i.d. hypothesis.

The normality assumption is also not strictly true.   Here we are saved by the fact that the money flux $\Delta_f$ is defined in terms of a logarithm, and while it has heavy tails, they are not sufficiently heavy to prevent it from converging to a normal.   We have explicitly verified this by tracking a group of funds in a given size range over time and demonstrating that normality is reached within 5 months.  Thus even though the normality assumption is not true on short timescales it rapidly becomes valid on longer timescales.

\begin{table}
\begin{center}
\small
\begin{tabular}{c|c|c|c|c|c}
\hline
date&12/2005&9/2005&6/2005&3/2005&12/2004\\
\hline\hline
$\beta_1$&$0.10\pm0.16$&$0.40\pm0.98$&$0.27\pm0.68$&$1.17\pm4.68$&$-0.23\pm1.24$\\
\hline
$\beta_2$&$0.14\pm0.27$&$0.36\pm1.20$&$0.48\pm0.83$&$-0.79\pm3.13$&$-0.65\pm2.31$\\
\hline
$\beta_3$&$0.28\pm0.45$&$0.01\pm1.07$&$0.33\pm0.83$&$1.79\pm3.24$&$0.60\pm2.57$\\
\hline
$\beta_4$&$0.56\pm0.40$&$-0.28\pm0.85$&$0.24\pm1.27$&$-0.28\pm1.65$&$0.44\pm2.32$\\
\hline
$\beta_5$&$0.24\pm0.43$&$-0.25\pm1.13$&$0.21\pm0.90$&$-0.24\pm2.95$&$0.43\pm2.49$\\
\hline
$\beta_6$&$0.48\pm0.38$&$-0.02\pm1.03$&$0.30\pm0.92$&$1.27\pm3.50$&$0.31\pm2.09$\\
\hline
$\beta$&$-0.02\pm0.02$&$0.03\pm0.05$&$0.01\pm0.05$&$0.14\pm0.21$&$0.06\pm0.15$\\
\hline
\end{tabular}\caption{\label{corr_table}
Cross-sectional regression coefficients of the monthly fund flow, computed for several months, against the performance in past months, as indicated in Eq.~\ref{performanceRegression}. The regression was computed cross-sectionally using data for 6189 equity funds. 
 For example the entry for $\beta_1$ in the first (from the left) column represents the linear regression coefficient of the money flux at the end of 2005 on the previous month's return.
 The errors are 95\% confidence intervals.}
\end{center}
\end{table}

\subsubsection{Size dependence of the growth process \label{sizeDependence}}

We now test the model for the size dependence of the growth process proposed in Section~\ref{muAndSigma}.  We also discuss the crucial role of the decomposition into returns and money flux in determining the size dependence.
\begin{center}
\begin{figure}
\includegraphics[width=12cm]{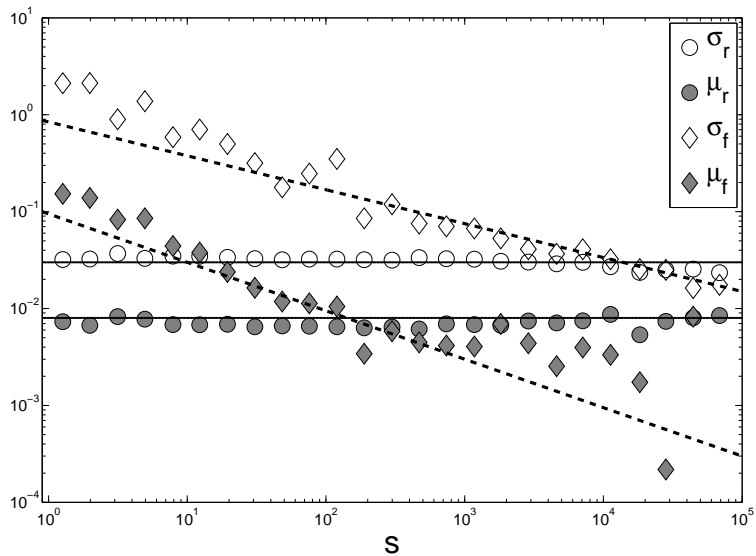}
\caption{\label{delta_vs_s}
A summary of the size dependence of mutual fund growth.  The average mean $\mu_r$ and volatility $\sigma_r$ of fund returns, as well as the average $\mu_f$ and volatility $\sigma_f$ of money flux (i.e. the flow of money in and out of funds), are plotted as a function of the fund size (in millions) for the year 2005 (see Eqs.~(\ref{ds} - \ref{df})).  The data are binned based on size, using bins with exponentially increasing size; we use monthly units. The average monthly return $\mu_r$ is compared to a constant return of 0.008 and the monthly volatility  $\sigma_r$ is compared to 0.03. The average monthly flux $\mu_f$ is compared to a line of slope of -0.5 and the money flux volatility $\sigma_f$ is compared to a line of slope -0.35. Thus absent any flow of money in or out of funds, performance is independent of size, as dictated by market efficiency.  In contrast, both the mean and the standard deviation of the money flows of funds decrease roughly as a power law as a function of size.}
\end{figure}
\end{center}
Figure~\ref{delta_vs_s} gives an overview of the size dependence for both the returns $\Delta_r$ and the money flux $\Delta_f$.  The two behave very differently.  The returns $\Delta_r$ are essentially independent of size\footnote{
The independence of the return $\Delta_r$ on size is verified by performing a linear regression of $\mu_r$ vs. $s$ for the year 2005, which results in an intercept $\beta=6.7\pm 0.2 \times10^{-3}$ and a slope $\alpha=0.5\pm 8.5 \times10^{-8}$. This result implies a size independent average monthly return of 0.67\%.%
  }.
This is expected based on market efficiency, as otherwise one could obtain superior performance simply by investing in larger or smaller funds \citep{malkiel-1995}.  This implies that equity mutual funds can be viewed as a constant return to scale industry \citep{gabaix-2006}.  Both the mean $\mu_r = E[\Delta_r]$ and the standard deviation $\sigma_r = \mbox{Var}[\Delta_r]^{1/2}$ are constant; the latter is also expected from market efficiency, as otherwise it would be possible to lower one's risk by simply investing in funds of a different size.  

In contrast, the money flux $\Delta_f$ decreases with size.  Both the mean money flux $\mu_f=E[\Delta_f]$ and its standard deviation $\sigma_f = \mbox{Var}[ \Delta_f]^{1/2}$ roughly follow a power law over five orders of magnitude in the size $s$.  This is similar to the behavior that has been observed for the growth rates of other types of firms \citep{stanley-1995,stanley-1996,Amaral97,Bottazzi03a}. As already discussed in footnote~9, there is a large body of theory attempting to explain this (and we believe our own theory presented elsewhere provides the correct explanation \citep{Schwarzkopf10b}).

As explained in Section~\ref{muAndSigma}, the steady state solution is qualitatively different depending on whether the parameters $\mu_\infty$ and $\sigma_\infty$ in Eq.~\ref{eq_mu_sigma_power_law} are positive.  As can be seen from the fit parameters in Table~\ref{table_fit}, based on data for $\Delta_s$ alone, we cannot strongly reject the hypothesis that the drift and diffusion rates vanish for large sizes, i.e. $\mu_{\infty}\to 0$ and $\sigma_{\infty} \to 0$. 
However, because the size change $\Delta_s$ can be decomposed as $\Delta_s = \Delta_r +\Delta_f$, efficiency dictates that $\Delta_r$ is independent of size, and since $E[\Delta_r] > 0$, we are confident that neither  $\mu_{\infty}$ nor $\sigma_{\infty}$ are zero.   
 
\begin{table}
\begin{center}
\begin{tabular}{c|c|c}
\hline 
variable  & 1991- 1998 &  1991- 2005 \\
\hline \hline
$\w_0$&$0.14$&$-0.37$\\
\hline
$\sigma_\w$&$3.02$&$3.16$ \\
 \hline\hline
$\sigma_0$&$0.35 \pm 0.02      $&$0.30 \pm 0.02$\\
\hline
$\beta$&$0.31 \pm 0.03$&$0.27 \pm 0.02$ \\
\hline
$\sigma_{\infty}$&$0.05 \pm 0.01      $&$0.05 \pm 0.01$\\
\hline
$R^2$&0.93&0.96\\
\hline\hline
$\mu_0$&$0.15 \pm 0.01    $&$0.08 \pm 0.05$\\
\hline
$\alpha$&$0.48 \pm 0.03      $&$0.52 \pm 0.04$\\
\hline
$\mu_{\infty}$&$0.002 \pm 0.008$&$0.004 \pm 0.001$ \\
\hline
$R^2$&0.98&0.97\\
\hline
\end{tabular}
\end{center}
\label{default}
\caption{\label{table_fit}
Model parameters as measured from the data in different time periods.  
$\w_0$ and $\sigma_\w^2$ are the mean and variance of the average (log) size of new funds described in (\ref{entry_distribution}).
$\sigma_0$, $\beta$ and $\sigma_\infty$ are the parameters for the size dependent diffusion and 
$\mu_0$, $\alpha$ and $\mu_{\infty}$ are the parameters of the average growth rate (\ref{eq_mu_sigma_power_law}). The confidence intervals are 95$\%$ under the assumption of standard errors.  The adjusted $R^2$ is given for the fits for each period.
The time intervals were chosen to match the results shown in Fig.~\ref{simV2}.
}
\end{table}

\begin{figure}
\begin{center}
\includegraphics[width=12cm]{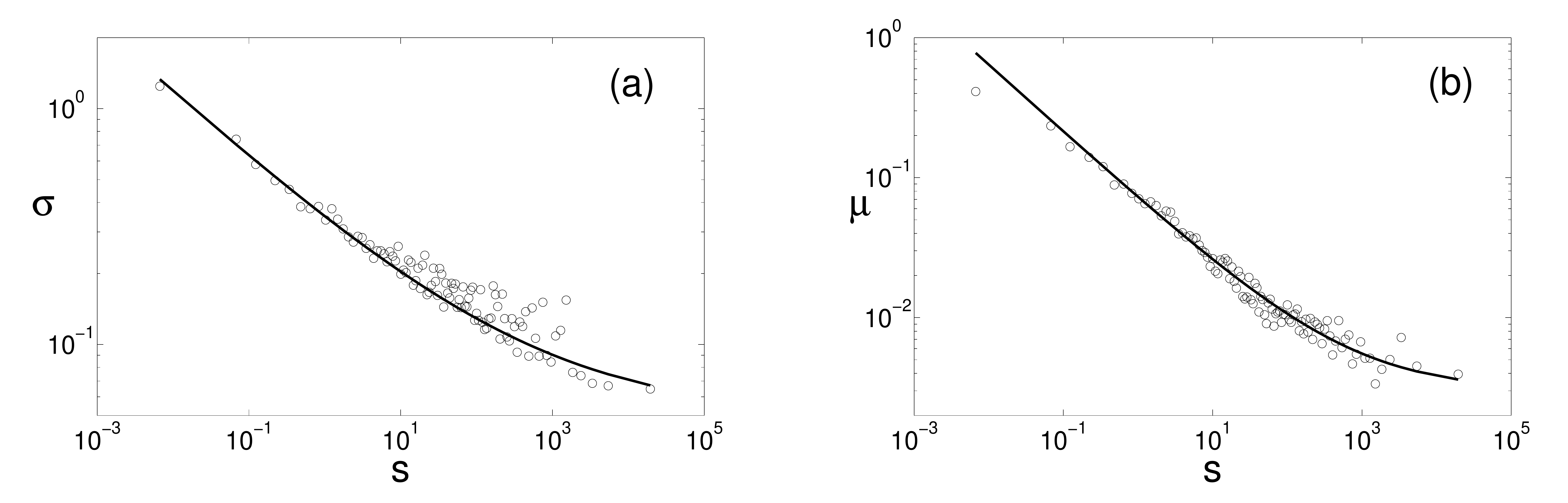}
\caption{\label{fit}
An illustration that the empirical power law-based model provides a good fit to the distribution of mutual funds.  (a) The standard deviation $\sigma$ of the logarithmic size change $\Delta_s = \Delta(\log s)$ of an equity fund as a function of the fund size $s$ (in millions of dollars). (b) The mean $\mu$ of $\Delta_s = \Delta(\log s)$ of an equity fund as a function of the fund size $s$ (in millions of dollars). The data for all the funds were divided into 100 equally occupied bins. $\mu$ is the mean in each bin and $\sigma$ is the square root of the variance in each bin for  the years 1991 to 2005.  The data are compared to a fit according to (\ref{eq_mu_sigma_power_law}) in Figures (a) and (b) respectively.  }
\end{center}
\end{figure}

As we showed in Table~\ref{corr_table}, the correlation between the returns $\Delta_r$ and the money flux $\Delta_f$ is small.   This implies that the standard deviations can be written as a simple sum.  Since $\Delta_r$ is independent of size and both the mean and standard deviation of $\Delta_f$ are power laws, this indicates that Eq.~\ref{muAndSigma} is a good approximation, and that $\mu_\infty$ and $\sigma_\infty$ are both greater than zero.
As illustrated in Figure~\ref{fit}, these functional forms fit the data reasonably well, with only slight variations of parameters in different periods, as shown in Table~\ref{table_fit}.  


\section{Testing the predictions of the model}\label{sec:comparison}

In this section we use our calibrated model of the entry, exit and size-dependent growth processes to simulate the evolution of the firm size distribution through time.  We are forced to use a simulation since,  once we include the size dependence of the diffusion and drift terms as given in equation~(\ref{eq_mu_sigma_power_law}), we are unable to find an analytic solution for the general diffusion equation (Eq.~(\ref{fokker-plank})). The analytic solution of the size independent case (Eq.~(\ref{time_dependent_solution})) gives the correct qualitative behavior, but the match is much better once one includes the size dependence.

The simulation was done on a monthly time scale, averaging over 1000 different runs to estimate the final distribution.  As we have emphasized in the previous discussion the time scales for relaxation to the steady state distribution are long.  It is therefore necessary to take the huge increase in the number of new funds seriously.  We begin the simulation in 1991 and simulate the process for varying periods of time, making our target the empirical distribution for fund size at the end of each period.   In each case we assume the size distribution for injecting funds is log-normal, as discussed in Section~\ref{sec:entry}.  

\begin{figure}
\begin{center}
\includegraphics[width=12cm]{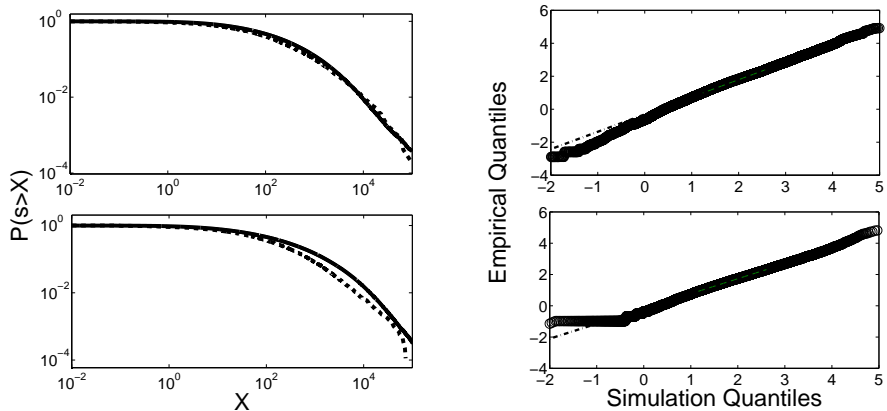}
\caption{
\label{simV2}
The model is compared to the empirical distribution at different time horizons.  The left column compares CDFs from the simulation (full line) to the empirical data (dashed line).  The right column is a QQ-plot comparing the two distributions.  In each case the simulation begins in 1991 and is based on the parameters in Table~\ref{table_fit}.  
The first row  corresponds to the years 1991-1998 and the second row to the years 1991-2005 (in each case we use the data at the end of the quoted year).
}
\end{center}
\end{figure}

To compare our predictions to the empirical data we measure the parameters for fund entry, exit and growth using data from the same period as the simulation, summarized in Table~\ref{table_fit}.  A key point is that we are not fitting these parameters on the target data for fund size\footnote{
It is not our intention to claim that the processes describing fund size are constant or even stationary.  Thus, we would not necessarily expect that parameters measured on periods outside of the sample period will be a good approximation for those in the sample period.  Rather, our purpose is to show that the random model for the entry, exit and growth processes can explain the distribution of fund sizes.},
but rather are fitting them on the individual entry, exit and diffusion processes and then simulating the corresponding model to predict fund size. One of our main predictions is that the time dependence of the solution is important.  In Figure~\ref{simV2} we compare the predictions of the simulation to the empirical data at two different ending times. 
The model fits quite well at all time horizons, though the fit in the tail is somewhat less good at the longest time horizon.  Note, that our simulations make it clear that the fluctuations in the tail are substantial.  The deviations between the prediction and the data are thus very likely due to chance -- many of the individual runs of the simulation deviate from the mean of the 1000 simulations more than the empirical data does.

\section{Conclusions\label{conclusions}}

We have argued that the mutual fund size distribution is driven by market efficiency, which gives rise to a random growth process.  The essential elements of the growth process are multiplicative random changes in the size of existing funds, entry of new funds, and exit of existing funds as they go out of business.  We find, however, that entry plays no role at all other than setting the scale; exit plays a small role in thickening the tails of the distribution, but this acts only on a very slow timescale.  The log-normality comes about because the industry is young and still in a transient state, and the exit process has not had a sufficient time to act.  In the future, if the conditions for fund growth and exit  were to remain stationary for more than a century, the distribution would become a power law.  The thickening of the tails happens from the body of the distribution outward, as the power law tail extends to successively larger funds.  We suspect that the conditions are highly unlikely to remain this stationary, and that the fund size distribution will remain indefinitely in its current log-normal, out of equilibrium state.

There is also an interesting size dependence in the growth rate of mutual fund size, which is both like and unlike that of other types of firms.  Mutual funds are distinctive in that their overall growth rates can be decomposed as a sum of two terms, $\Delta_s = \Delta_f + \Delta_r$, where $\Delta_f$ represents the flow of money in and out of funds, and $\Delta_r$ the returns on money that is already in the fund.  The money flow $\Delta_f$ decreases as a power law as a function of size, similar to what is widely observed in the overall growth rates for other types of firms.  Furthermore the exponents are similar to those observed elsewhere.  The returns $\Delta_r$, in contrast, are essentially independent of fund size, as they must be under market efficiency.  As a result, for large sizes the mean and variance of the overall growth are constant -- this is unlike other firms, for which the mean and variance appear to go to zero in the limit.  As we discuss here, this makes a difference in the long-term evolution:  While the exit process is driving mutual funds to evolve toward a heavier-tailed distribution, other firms are evolving toward a thinner-tailed distribution.  Again, though, due to the extremely slow relaxation times, we suspect this makes little or no difference.

Our analysis here suggests that the details of investor preference have a negligible influence on the upper tail of the mutual fund size distribution, except insofar as investors choose funds so as to enforce market efficiency.  Investor preference enters our analysis only through $\Delta_f$, the flow of money in and out of the fund.  Since $\Delta_f$ becomes relatively small in the large size limit, the growth of large funds is dominated by the returns $\Delta_r$, whose mean and variance are constant.  Thus the upper tail of the size distribution is determined by market efficiency, which dictates both that returns are essentially random, and thus diffusive, and that there is no dependence on size.  As a result,  for large fund size investor preference doesn't have much influence on the growth process.  This is reinforced by the fact that the statistical properties of the money flux $\Delta_f$ are essentially like those of the growth of other firms.  

How can size-dependent transaction costs be compatible with our results here?  We have performed an empirical study, which we will report elsewhere,  that demonstrates that as size increases fund managers maintain constant after-transaction cost performance by lowering fees, reducing trading and diversifying investments.  This is in contrast to the theory proposed by Berk and Green (\citeyear{Berk04}) that fund size is determined by the skill of fund managers, i.e. that better managers attract more investment until increased transaction cost causes excess returns to disappear.  Both our theory and that of Berk and Green are based on market efficiency.  The key difference is that we find that the flatness of performance vs. size is enforced by simple actions taken by fund managers that do not influence the diffusion of fund size.   In contrast, the Berk and Green theory requires choices by investors that directly influence fund size, and thus is not compatible with the free diffusion that we have prevented empirical evidence for here.  In their theory transaction costs and investor skill determine fund size; in our theory, neither plays a role.

We would like to stress that, while we are fitting econometric models to the entry, exit and growth processes, and calibrating these models against the data, we are not fitting any parameters on the size data itself.  This makes it challenging to get a model that fits as well as the model shown in Fig.~\ref{simV2}.  Of course, we have only demonstrated that the random process model is sufficient to explain fund size; we cannot demonstrate that other explanations might not also be able to explain it.  However, the assumptions that we make here are simple and natural.  The stochastic nature of fund growth is not surprising:  It is well known that past returns do not predict future returns.  Thus even if investors chase returns, they are chasing something that is inherently random.  We believe that this is at the core of why our model works so well.  Our demonstration that a good explanation can be obtained based on market efficiency alone, which requires weaker assumptions than full rationality, provides a theory that is robust and largely independent of the details of human choice.
 
%

 \section*{Acknowledgements}

{\small We would like to thank Brad Barber, Giovani Dosi, Fabrizio Lilo, Terry Odean, Eric Smith and especially Rob Axtell and Andrew Lo for useful comments. YS would like to thank Mark B. Wise. We gratefully acknowledge financial support from NSF grant HSD-0624351.  Any opinions, findings and conclusions or recommendations expressed in this material are those of the authors and do not necessarily reflect the views of the National Science Foundation.}

\appendix
\newpage
\section{Inadequacy of Gini coefficients to characterize tail behavior}\label{gini}

\begin{figure}[htbp]
\begin{center}
\includegraphics[width=6cm]{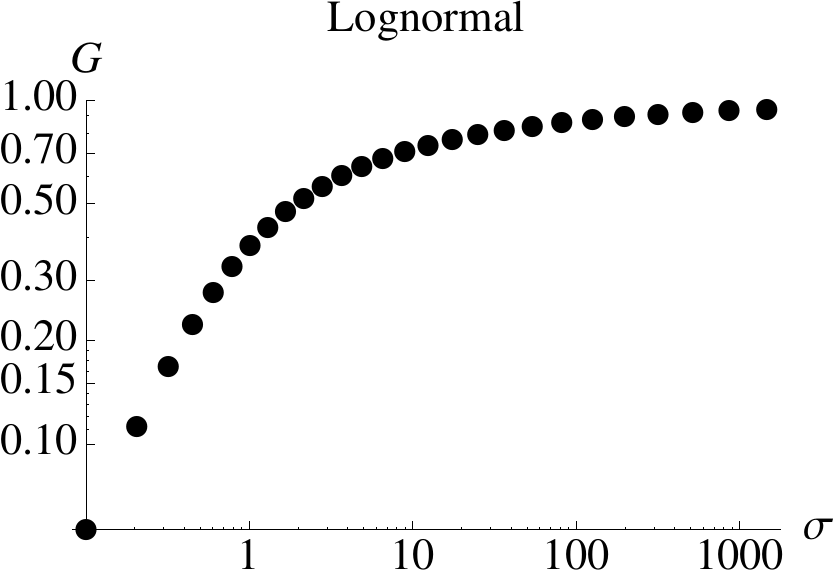}
\includegraphics[width=6cm]{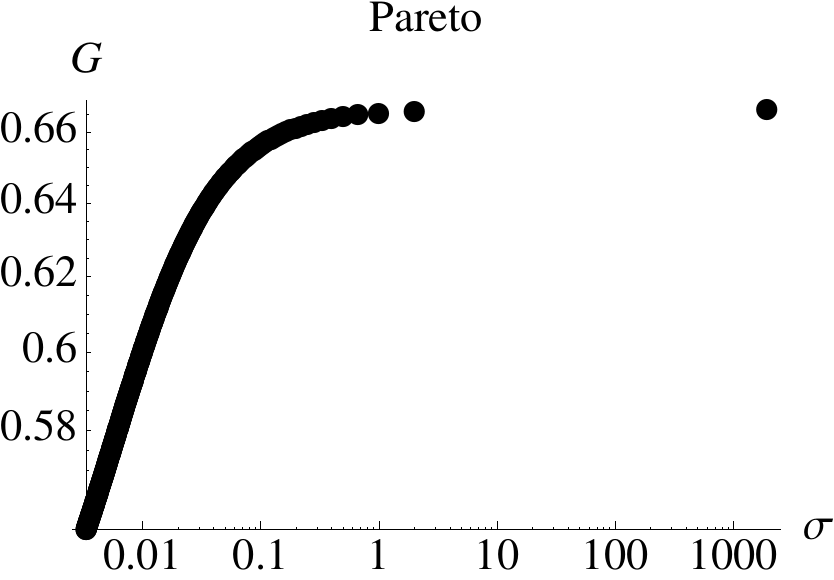}
\caption{The Gini coefficients as described in equation (\ref{eq:gini}) are calculated numerically for a lognormal distribution and a Pareto distribution. The Gini Coefficients were calculated for different parameter values and are plotted as a function of the resulting standard deviation. For the Pareto distribution (footnote 20) we used $s_0=0.01$ and different exponents $\alpha$ in the range $(2,5]$, i.e. a finite second moment. The lower standard deviation $\sigma=0.0033$ corresponds to $\alpha=5$ and $\sigma=1916.17$ corresponds to $\alpha\to2$. For the lognormal we used $a=0$ and different $b$ in the range $[0.1,2.8]$ where $b=0.1$ corresponds to $\sigma=0.101$ and $b=2.757$ corresponds to $\sigma=2000$. 
}
\label{fig:gini}
\end{center}
\end{figure}

The Gini coefficient \citep{Gini1912} is commonly used as a measure of inequality 
but as we show here it is not suitable for distinguishing between highly skewed distributions when one wishes to focus on tail behavior.
For a non negative size $s$ with a CDF $F(s)$,  the Gini coefficient $G$ is given by 
\begin{equation}\label{eq:gini}
G=\frac{1}{E[s]}\int_0^{\infty} F(s)(1-F(s))\mathrm{d}s,
\end{equation}
where $E[s]$ is the mean \citep{Dorfman79}.
To illustrate the problem we compare the Gini coefficients of a Pareto distribution to those of a lognormal\footnote{
The CDF of the Pareto distribution is defined as 
\begin{equation}\label{eq:cdf_pareto}
F_{p}(s)=1-\left(\frac{s}{s_0}\right)^{-\alpha},
\end{equation}
where $s_0$ is the minimum size and $\alpha$ is the tail exponent.
The CDF of a lognormal is given by 
\begin{equation}\label{eq:cdf_ln}
F_{ln}(s)=\frac{1}{2} \left(1+\mathrm{Erf}\left[\frac{\log(s)-a}{\sqrt{2} b}\right]\right),
\end{equation}
where $a$ is a location parameter and $b$ is the scale parameter.}.
For a Pareto distribution with tail parameter $\alpha$ the $m>\alpha$ moments do not exist.  This is in contrast to the lognormal distribution, for which all moments exist.   Naively one would therefore expect that the Gini coefficient of the Pareto distribution (see footnote 18) to be larger than that of a lognormal since it has a heavier upper tail.   This is true for a Pareto distribution with $\alpha < 2$, for which the Gini coefficient  is one due to the fact that the standard deviation does not exist.  However, when $\alpha < 2$, for large standard deviations the Gini coefficient of the log-normal is greater than that of the Pareto, as shown in Figure~\ref{fig:gini}.  In order to compare apples to apples in Figure~\ref{fig:gini} we plot the Gini coefficient as a function of the standard deviation  (which is a function of the distribution parameters). For a Pareto distribution with a finite second moment ($\alpha>2$) the lognormal has a higher coefficient.

Thus, even though the Gini coefficient is frequently used as measure for inequality, it is not a good measure when one seeks to study tail properties, particularly for comparisons of distributions with different functional representations.  The reason is that the Gini coefficient is a property of the whole distribution, and depends on the shape of the body as well as the tail.  Similar remarks apply to the Herfindahl index.

\section{Simulation model}\label{appendix_simulation_model}
We simulate a model with three independent stochastic processes. These processes are modeled as
Poisson process and as such are modeled as having at each time step a probability for an event to occur.
The simulation uses asynchronous updating to mimic continuous time.
At each simulation time step we 
 perform one of three events with an appropriate probability.
 These probabilities  will determine the rates in which that process occurs.  
 The probability ratio between any pair of events  should be equal to the ratio of the rates of the corresponding processes.
 Thus, if we want to simulate this model for given rates our probabilities are determined.  
 
 These processes we simulate are:
\begin{enumerate}\label{rate_description}
\item The rate of size change taken to be 1 for each fund and $N$ for the entire population.\newline
Thus, each fund changes size with a rate taken to be unity.
\item The fund exit rate $\lambda$ which can depend on the fund size.
\item The rate of creation of new funds $\nu$.\newline
Each new fund enters with a size $\w$ with a probability density $f(\w)$.    
\end{enumerate}

 Since some of these processes are defined per firm as opposed to the creation process, the simulation is not straightforward. We offer a brief description of our simulation procedure.
\begin{enumerate}
\item At every simulation time step, with a probability $\frac{\nu}{1+\lambda +\nu}$  a new fund enters and we proceed to the next simulation time step. 
\item If a fund did not enter then the following is repeated $(1+\lambda)N$ times. 
\begin{description}
\item[a.] We pick a fund at random.  
\item[b]. With a probability of $\frac{\lambda}{1+\lambda}$ the fund enters.
\item[c.] If it is not annihilated, which happens with a probability of   $\frac{1}{1+\lambda}$, we change the fund size. 
\end{description}
\end{enumerate}

We are interested in comparing the simulations to both numerical and empirical results.
The comparisons with analytical results are done for specific times and for specific years when comparing to  empirical data.
In order to do so, we need to convert simulation time to "real" time.
 The simulation time can be compared to 'real' time if every time a fund does not enter
we add a time step. Because of the way we defined the probabilities each simulation time step is comparable to $1/(1+\lambda)$ in "real" time units.
 The resulting "real" time is then measured in what ever units our rates were measured in.
 In our simulation we use monthly rates and as such a unit time step corresponds to one month.


\end{document}